\documentclass[a4paper, 12pt]{article}
\usepackage{amsmath}
\usepackage{amsthm}
\usepackage{amssymb}
\usepackage{graphicx}
\usepackage{enumerate}
\usepackage{natbib}
\usepackage{url} 
\usepackage{comment}
\usepackage{booktabs}
\usepackage[toc]{appendix}

\usepackage{color}
\usepackage{xcolor, soul}
\usepackage{multirow}
\usepackage{longtable}
\let\oldproofname=\proofname
\renewcommand{\proofname}{\rm\bf{\oldproofname}}
\usepackage{comment}
\usepackage{placeins}

\usepackage{float}
\usepackage{hyperref}
\usepackage{float}
\hypersetup{colorlinks,
    citecolor=black,
    filecolor=black,
    linkcolor=black,
    urlcolor=black,
    breaklinks=true}

\numberwithin{equation}{section}
\usepackage{titlesec}

\titleformat{\section}
  {\normalfont\Large\bfseries}
  {\thesection.}{1em}{}

\newcommand{\blind}{1}

\def \ba {{\mathbf a}}

\def \be {{\mathbf e}}

\def \bu {{\mathbf u}}

\def \bx {{\mathbf x}}
\def \by {{\mathbf y}}

\def \bA {{\mathbf A}}

\def \bG {{\mathbf G}}

\def \bI {{\mathbf I}}

\def \bV {{\mathbf V}}

\def \bX {{\mathbf X}}
\def \bY {{\mathbf Y}}
\def \bZ {{\mathbf Z}}

\def \bmu {\boldsymbol{\mu}}

\def \bbeta {\boldsymbol{\beta}}

\def \btheta {\boldsymbol{\theta}}

\def \stackind {\stackrel{{ind}}{\sim}}
\def \stackiid {\stackrel{{iid}}{\sim}}

\def \bSigma {\boldsymbol{\Sigma}}
\def \bGamma {\boldsymbol{\Gamma}}

\def \bPhi {\boldsymbol{\Phi}}

\def\cero{\mathbf 0}

\def\var{\mbox{Var}}

\date{}
\usepackage{authblk} 
\usepackage{ragged2e}

\usepackage{setspace}

\begin{document}

\doublespacing

\if1\blind
{
  \title{\bf Pseudo Empirical Best Prediction of Multiple Characteristics in Small Areas}
  \author[1]{William Acero\thanks{Supported by the grant PID2020-115598RB-I0 funded by the Ministry of Science and Innovation (Spain).}}
  \author[2]{Domingo Morales \thanks{Supported by the CIPROM/2024/34 grant funded by the Conselleria de Educación, Cultura, Universidades y Empleo, Generalitat Valenciana (Spain) and by the grant PID2022-136878NB-I00 funded by MCIN/AEI/10.13039/501100011033 (Spain)}}
  \author[1,3]{Isabel Molina}
  \affil[1]{Department of Statistics and Operation Research, Complutense University of Madrid, Madrid, Spain}
  \affil[2]{Operations Research Center, Miguel Hernández University of Elche, Elche, Spain}
  \affil[3]{Interdisciplinary Mathematics Institute (IMI), Complutense University of Madrid, Madrid, Spain}
  \maketitle
} \fi

\if0\blind
{
  \bigskip
  \bigskip
  \bigskip
  \begin{center}
    {\LARGE\bf Pseudo Empirical Best Prediction of Multiple Characteristics in Small Areas }
\end{center}
\medskip
} \fi

\begin{singlespace}
\begin{abstract}
\normalsize
Small area estimators that ignore the sampling design lack design consistency when the sampling mechanism is complex and may be severely biased under informative designs. Existing procedures that account for the survey weights under unit-level models typically focus on a single response variable. This paper addresses the estimation of area means for several dependent target variables under a multivariate nested error regression (MNER) model. We propose a multivariate pseudo–empirical best linear unbiased predictor that accounts for the sampling mechanism. Moreover, by aggregating the MNER model, we derive a unified predictor that can be obtained from either unit-level or area-level data. Bootstrap procedures are proposed to estimate the mean squared errors (MSEs) of the proposed predictors. Simulation experiments are conducted to examine the properties of the proposed small area estimators and the MSE estimators. Finally, an application with housing data illustrates the proposed methods.
\end{abstract}

\noindent%
\end{singlespace}

\section{Introduction}

When estimating means for domains with small sample sizes, traditional design-based direct estimators are unreliable. Small area estimation (SAE) methods, as described by \cite{rao2015small} and \cite{morales2021course}, improve the precision of direct estimators by borrowing strength over areas. Among SAE procedures, model-based approaches based on unit-level data can produce substantial efficiency gains. In the frequentist setup, probably the most popular unit-level procedure is the empirical best linear unbiased predictor (EBLUP) proposed by \cite{battese1988error} under a nested error regression (NER) model. Under a hierarchical Bayes (HB) approach for SAE, we can also find many approaches, such as those of \cite{datta1991bayesian}, \cite{ghosh1992hierarchical} and \cite{you2000hierarchical}.
None of these methods explicitly incorporate the survey weights. As a consequence, the resulting small area estimators lack design consistency as the area sample sizes increase, unless the sampling design is self-weighting within the areas \citep{you2002pseudo}. 

On the other hand, the EBLUP based on the area-level Fay–Herriot (FH) model proposed by \cite{fay1979estimates} accounts for the survey design, but it is likely to produce estimators with lower efficiency than the alternative ones obtained by unit-level procedures, specially when the number of areas $D$ is not so large. Another drawback of the FH model is that the error variances are assumed to be known for all areas. In practice, these variances are previously estimated using the available unit-level data from each area. However, the additional uncertainty of the EBLUP due to the estimation of the error variances is often ignored in the mean squared error (MSE).

When estimating the area means of a single response variable, several model-based estimators that incorporate the survey weights have been proposed in the literature. \cite{prasad1999robust} introduced the Pseudo-EBLUP (PEBLUP) under the NER model, based on taking weighted means in the NER model over the sampled units in each area, using the survey weights. The resulting aggregated area-level model is fitted using the corresponding area-level data. Later, \cite{you2002pseudo} noted that a more efficient PEBLUP can be obtained using the available unit-level survey data to fit the original unit-level NER model. Using the same ideas, \cite{YOU2003197} proposed Pseudo-Hierarchical Bayes small area estimators. 

In the aggregated area-level model introduced by \cite{prasad1999robust} and \cite{you2002pseudo}, the error variances depend on a common unknown variance parameter for all areas, which can be consistently estimated based on the overall survey data. Hence, the PEBLUP avoids the assumption of known error variances and MSE estimators that incorporate the uncertainty from the error variance estimators may be obtained.

Consider now that the aggregation of the NER model as in \cite{you2002pseudo} is based on calibrated sampling weights, such that the expansion estimators of the area totals of the auxiliary variables in the NER model are equal to the true area totals, as in \cite{Acero2025}. Then, the result is a kind of FH model, with a more convenient specification of error variances in terms of a single model parameter. Accordingly, the resulting PEBLUP equals the usual EBLUP under that modified FH model and is named ``unified'' predictor by \cite{Acero2025}. As in the PEBLUP, the unknown parameters in the aggregated model can also be estimated using the available unit-level survey data, which may lead to a more efficient unified predictor than when using only the corresponding area-level aggregates.  

Multivariate area- or unit-level mixed models have been considered when estimating small area parameters for several dependent response variables. \cite{fay1987application} and \cite{datta1991} showed better performance of the small area estimators obtained from multivariate models compared to those obtained from separate univariate models. \cite{BENAVENT2016372} proposed a multivariate FH (MFH) model for SAE. In this model, the error covariance matrices are assumed to be known and equal to the corresponding design-based covariance matrices of the vectors of direct estimators, for all the areas. However, as in the univariate FH model, in practice the error covariance matrices are often replaced by design-based estimators, whose uncertainty is typically ignored. \cite{esteban2022small} obtained the EBLUPs of a bivariate area mean under a bivariate NER (BNER) model, and considered a plug-in predictor for domain ratios. Design-consistent small area estimators of multiple characteristics obtained from multivariate unit-level models are still lacking in the literature.

In this paper, we are interested in the design-consistent estimation of area means for several dependent response variables. For this purpose, we extend the PEBLUP of \cite{you2002pseudo} by considering a multivariate nested error (MNER) model for the vector of response variables. The PEBLUP is obtained by aggregating the MNER model, yielding an aggregated multivariate area-level model that incorporates the survey weights, and whose error covariance matrices depend on an unknown matrix that is common for all the areas. 

We illustrate how, when considering calibrated weights at the totals of the covariates for each area, this PEBLUP becomes a multivariate ``unified" predictor, which may be obtained using either the unit-level data or the area-level data, and whose error covariance matrices may be estimated consistently (as the number of areas increases) combining the data from all areas.

The paper is organized as follows. We start describing the notation and the target parameters in Section \ref{secNotation}, and the assumed MNER model in Section \ref{SectionMNER}. In Section \ref{sec:MPEBLUP}, we introduce the proposed multivariate versions of the Pseudo-EBLUP and the unified predictor. In Section \ref{sec:mse}, we study the estimation of the mean squared error (MSE) matrix. Section \ref{sec:sim} evaluates the performance of the proposed predictors and the bootstrap MSE estimators through simulation experiments. In Section \ref{application}, we illustrate the application of the proposed predictors in the context of housing price and rent data from Colombia. Finally, concluding remarks are provided in Section \ref{sec:conclusions}.

\section{Notation and target parameters}\label{secNotation}

We consider a finite population $U$ of size $N$, composed of non-overlapping domains $U_1, \dots, U_D$, of sizes $N_1,\ldots,N_D$, with $N=\sum_{d=1}^D N_d$. We denote by $\by_{di}=(y_{di1},\ldots,y_{diR})^t$ the vector with the measurements of $R$ dependent continuous target variables for the $i$th unit from domain $d$, for $i=1,\ldots,N_d$ and $d=1,\ldots,D$. The dimension $R$ does not depend on the number of areas $D$. The aim is to estimate the domain means of the $R$ target variables, 
$$
\bmu_d=(\mu_{d1}, \ldots, \mu_{dR})^t = N_d^{-1}\sum_{i=1}^{N_d} \by_{di},\quad d=1\ldots,D.
$$

To estimate $\bmu_d$, we consider a sample $s_d$ of size $n_d=|s_d|<N_d$ drawn from domain $U_d$, with $n_d>0$, $d=1,\ldots,D$. Let $n=\sum_{d=1}^D n_d>D$ be the total size of the overall sample $s=s_1\cup\cdots\cup s_D$. We denote by $\pi_{di}>0$ the inclusion probability of unit $i$ in the sample $s_d$ from area $d$, $w_{di}=\pi_{di}^{-1}$ the corresponding sampling weight, and $w_{d\cdot}=\sum_{i \in s_d} w_{di}$, $d=1,\ldots,D$. Based on the area-specific unit-level survey data $\{(\by_{di}, w_{di}), \, i \in s_d\}$, we can obtain a vector $\hat{\bmu}_d^{DIR}=(\hat{\mu}_{d1}^{DIR},\ldots,\hat{\mu}_{dR}^{DIR})^t$ of direct estimators, such as the H\'ajek estimators
\begin{align}\label{Hajek}
\hat{\bmu}_d^{DIR}=\bar{\by}_{dw} = w_{d\cdot}^{-1}\sum_{i \in s_d} w_{di} \by_{di}, \quad d=1,\ldots,D.
\end{align} 
The inefficiency of direct estimators for domains with a small sample size $n_d$ calls for the application of model-based small area estimation procedures discussed next.

\section{Multivariate nested-error linear regression model}\label{SectionMNER}

Consider that, from other data sources (census, administrative records,...), we have vectors ${\bx}_{dir}=(x_{dir1},\ldots,x_{dirp_r})^t$ with the values of $p_r$ explanatory variables associated with the $r$th response variable $y_{dir}$, $r=1,\ldots,R$. We denote by $p=\sum_{r=1}^Rp_r$ the total number of covariates for the $R$ target variables and define the $R\times p$ matrix ${\bX}_{di}=\mbox{diag}(\bx_{di1}^t, \ldots, \bx_{diR}^t)$.
We assume that the population units follow the MNER model given by
\begin{equation}\label{MNERdi}
\by_{di}=\bX_{di}\bbeta+\bu_d+\be_{di},\quad i=1,\ldots,N_d,\ d=1,\ldots,D.
\end{equation}
In this model, $\bbeta=(\bbeta_{1}^t ,\ldots,\bbeta_{R}^t)^t$, where
$\bbeta_{r}$ is a $p_r\times 1$ vector of regression coefficients for the $r$th vector of covariates $\bx_{dir}$, $r=1,\ldots,R$. The vectors of domain effects $\bu_{d}$ and unit-level errors $\be_{di}$, $i=1,\ldots,N_d$, $d=1,\ldots,D$, are all mutually independent, satisfying
\begin{eqnarray}
&& \bu_{d}=(u_{d1}, \ldots, u_{dR})^t \stackrel{iid}\sim N_R(\cero_R,\bSigma_u(\btheta)),\nonumber\\
&& \be_{di}=(e_{di1}, \ldots, e_{diR})^t \stackrel{iid}\sim N_R(\cero_R,\bSigma_e(\btheta)), \quad i=1,\ldots,N_d, \ d=1,\ldots,D.\label{MNERdi2}
\end{eqnarray}
Here, $\bSigma_u(\btheta)=(\sigma_{u,r\ell})_{r,\ell=1,\ldots,R}$ and $\bSigma_e(\btheta)=(\sigma_{e,r\ell})_{r,\ell=1,\ldots,R}$ are general positive definite symmetric matrices of unknown elements, satisfying $\sigma_{ur}^2:=\sigma_{u,rr}\in (0,\infty)$ and $\sigma_{er}^2:=\sigma_{e,rr}\in (0,\infty)$, $r=1,\ldots,R$, $\sigma_{u,r\ell}=\sigma_{u,\ell r}\in (-\infty,\infty)$ and $\sigma_{e,r\ell}=\sigma_{e,\ell r}\in (-\infty,\infty)$, for $r\neq \ell$.
The vector with all the unknown variance components is
$$
\btheta =(\sigma_{u,1}^2,\ldots,\sigma_{u,R}^2,\sigma_{u,12},\ldots,\sigma_{u,R-1,R},\sigma_{e,1}^2,\ldots,\sigma_{e,R}^2,\sigma_{e,12},\ldots,\sigma_{e,R-1,R})^t.
$$ 
For the case of $R=2$ response variables, model \eqref{MNERdi}-\eqref{MNERdi2} reduces to the BNER model studied by \cite{esteban2022small}. 

The model \eqref{MNERdi}-\eqref{MNERdi2} is fitted to the unit-level survey data $\{(\by_{di}, \bX_{di}), \ i\in s_d, \ d=1,\ldots,D\}$. 
When the sample design is ignorable (see e.g, \citealp{pfeffermann2007small}, \citealp{cho2024optimal}), the sample units $i\in s_d$, $d=1,\ldots,D$, follow the same model as the population units. Let us define the area vectors and matrices for the sample units,
$$
\by_{ds}=\underset{i\in s_d}{\mbox{col}}(\by_{di}),\
\bX_{ds}=\underset{i\in s_d}{\mbox{col}}(\bX_{di}),\ 
\bZ_{ds}=\underset{i\in s_d}{\mbox{col}}(\bI_R),\
\be_{ds}=\underset{i\in s_d}{\mbox{col}}(\be_{di}),\
\bV_{eds}(\btheta)=\underset{i \in s_d}{\mbox{diag}}(\bSigma_e(\btheta)).
$$
In this notation, the MNER model for the sample units is
\begin{equation}\label{mod2NERd}
\by_{ds}=\bX_{ds}\bbeta+\bZ_{ds}\bu_d+\be_{ds},\quad d=1,\ldots,D,
\end{equation}
where $\be_{ds}\stackind N_{Rn_d}(\cero_{Rn_d},\bV_{eds}(\btheta))$ and $\bu_d$, $d=1,\ldots, D$, are
 mutually independent. The covariance matrix of $\by_{ds}$ is given by 
$$
\bV_{ds}(\btheta)=\bZ_{ds}\bSigma_u(\btheta)\bZ_{ds}^t + \bV_{eds}(\btheta).
$$
For known $\btheta$, we can estimate $\bbeta$ by the WLS estimator, equal to the maximum likelihood (ML) estimator under the MNER model \eqref{MNERdi}--\eqref{MNERdi2}, and given by
\begin{equation}\label{WLS}
\tilde{\bbeta}(\btheta)= 
\left(\sum_{d=1}^D \bX_{ds}^t \bV_{ds}^{-1}(\btheta)\bX_{ds}\right)^{-1}\sum_{d=1}^D \bX_{ds}^t \bV_{ds}^{-1}(\btheta)\by_{ds}.
\end{equation}
In addition, a consistent estimator $\hat{\btheta}$ of $\btheta$ as $D\to\infty$ can be obtained e.g. by maximum likelihood (ML) or restricted/residual ML (REML). Replacing $\hat{\btheta}$ for $\btheta$ in \eqref{WLS}, we obtain the final estimator of $\bbeta$, denoted $\hat{\bbeta}=\tilde{\bbeta}(\hat{\btheta})$. 
Details of the REML fitting procedure for estimating $\btheta$ under the MNER model with $R=2$ are provided in \cite{esteban2022small}. This fitting procedure can automatically be extended to $R>2$.

\section{Multivariate Pseudo-EBLUP}\label{sec:MPEBLUP}

Take a weighted average over the sample units $i\in s_d$ from area $d$ in the MNER model \eqref{MNERdi}-\eqref{MNERdi2}, using the sampling weights $w_{di}$, $i\in s_d$. The result is the area-level model
\begin{equation}\label{FHw}
\bar{\by}_{dw}=\bar{\bX}_{dw} \bbeta+\bu_d+\bar{\be}_{dw},\quad d=1,\ldots,D,
\end{equation}
where $\bar{\bX}_{dw}=w_{d\cdot}^{-1}\sum_{i\in s_d}w_{di}\bX_{di}$ and
$\bar{\be}_{dw}=w_{d\cdot}^{-1}\sum_{i\in s_d}w_{di}\be_{di}$. Note that $\bar{\be}_{dw}$ satisfies
$$
\bar{\be}_{dw} \stackind N_R(\cero_R,\bSigma_{edw}(\btheta)),\quad \bSigma_{edw}(\btheta)=k_d^2\,\bSigma_e(\btheta)
$$
for $k_d^2=w_{d\cdot}^{-2}\sum_{i\in s_d}w_{di}^2$, $d=1,\ldots, D$.
Under the aggregated model \eqref{FHw}, the covariance matrix of $\bar{\by}_{dw}$ is given by
\begin{equation}\label{Vdw}
\bV_{dw}(\btheta)=\bSigma_u(\btheta)+\bSigma_{edw}(\btheta),\quad d=1,\ldots,D.
\end{equation}

Let us define the $R\times p$ matrix $\bar{\bX}_d=N_d^{-1}\sum_{i=1}^{N_d}\bX_{di}$, $d=1,\ldots,D$.
Under the MNER model \eqref{MNERdi}--\eqref{MNERdi2}, by the Strong Law of Large Numbers, for $N_d$ large, the vector of target area means is $\bmu_d\approx \bar{\bX}_d \bbeta + \bu_d$, $d=1,\ldots,D$. Since domain population sizes $N_d$ are typically very large, from now on, we write equality in this approximation.

In terms of the vector of target area means $\bmu_d=\bar{\bX}_d\bbeta+\bu_d$, we define the univariate domain parameter $\delta_d=\ba^t\bmu_d$, where $\ba=(a_1,\ldots,a_R)^t$ is a known non-stochastic vector. 
The Best Predictor (BP) of $\delta_d=\ba^t\bmu_d$ based on the aggregated model \eqref{FHw}
 minimizes the MSE under that model and is given by $\tilde\delta_d(\btheta,\bbeta)=\ba^t\tilde\bmu_d(\btheta,\bbeta)$, where $\tilde\bmu_d(\btheta,\bbeta)=E(\bmu_d|\bar{\by}_{dw})$. In this paper, we refer to $\tilde\bmu_d(\btheta,\bbeta)$ as the Multivariate Pseudo-Best Predictor (MPBP) of $\bmu_d$. 
Note that the joint distribution of $\bmu_d$ and $\bar{\by}_{dw}$ is given by
$$
\left(\begin{array}{l}
\bar{\by}_{dw}\\
\bmu_d
\end{array}\right) \sim N\left(\left(\begin{array}{l}
\bar{\bX}_{dw}\bbeta\\
\bar{\bX}_d \bbeta
\end{array}\right),
\left(\begin{array}{ll}
\bV_{dw}(\btheta) & \bSigma_u(\btheta)\\
\bSigma_u(\btheta) & \bSigma_u(\btheta)
\end{array}\right)\right).
$$
The MPBP of $\bmu_d$ is the conditional expectation $E(\bmu_d|\bar{\by}_{dw})$ obtained from this joint distribution. Denoting $\bGamma_{dw}(\btheta) = \bSigma_u(\btheta)\bV_{dw}^{-1}(\btheta)$, the MPBP of $\bmu_d$ reads
\begin{eqnarray}
\tilde{\bmu}_d(\btheta,\bbeta) &=& 
\bar{\bX}_d \bbeta +\bGamma_{dw}(\btheta)\left(\bar{\by}_{dw}-\bar{\bX}_{dw}\bbeta\right)\nonumber\\
&=&\bGamma_{dw}(\btheta)\big[\bar{\by}_{dw}+(\bar{\bX}_d-\bar{\bX}_{dw})\bbeta\big]+(\bI_R-\bGamma_{dw}(\btheta))\bar{\bX}_d \bbeta,\label{YRpseudoblup}
\end{eqnarray}

Similarly as in \cite{you2002pseudo}, $k_d^2\to 0$ as $n_d\to\infty$, provided that $w_{di}>0$, $\forall i \in s_d$ and $\max_{i\in s_d}(w_{di}/w_{d\cdot})=O(n_d^{-1})$. Since the elements of $\bSigma_e(\btheta)$ are bounded, then $\bV_{dw}(\btheta)\to \bSigma_u(\btheta)$, which leads to $\bGamma_{dw}(\btheta) \to \bI_R$ as $n_d \to \infty$. The latter result implies design consistency of the Pseudo BP $\tilde\delta_d(\btheta,\bbeta)=\ba^t\tilde\bmu_d(\btheta,\bbeta)$ for $\delta_d=\ba^t\bmu_d$, under the same conditions under which $\ba^t\bar\by_{dw}$ is design consistent for $\delta_d=\ba^t\bmu_d$, as $n_d\to \infty$.

Note that, replacing the unknown $\bbeta$ by the WLS estimator based on the aggregated model \eqref{FHw} in the Pseudo BP $\tilde\delta_d=\ba_d^t\tilde{\bmu}_d(\btheta,\bbeta)$, with $\tilde{\bmu}_d(\btheta,\bbeta)$ given in \eqref{YRpseudoblup}, we obtain the BLUP of $\delta_d=\ba_d^t\bmu_d$ under that model. Instead, we use the unit-level data to estimate $\bbeta$. In order to account for the design, we use a multivariate extension of the estimator of $\bbeta$ by \cite{you2002pseudo}. Denoting $\tilde\bu_{dw}(\bbeta,\btheta)=\bGamma_{dw}(\btheta)\left(\bar{\by}_{dw}-\bar{\bX}_{dw}\bbeta\right)$, the extended estimator of $\bbeta$ is the solution to the following survey-weighted estimating equation
\begin{equation}\label{sweq}
\sum_{d=1}^{D}\sum_{i\in s_d}w_{di}\bX_{di}^t \left(\by_{di}-\bX_{di}\bbeta-\tilde\bu_{dw}(\btheta,\bbeta)\right)=\cero_R.
\end{equation}
The solution reads 
\begin{equation}\label{betaw}
    \tilde\bbeta_{w}(\btheta) = \left[\sum_{d=1}^{D}\sum_{i\in s_d} w_{di}\bX_{di}^t \left(\bX_{di}- \bGamma_{dw}(\btheta) \bar{\bX}_{dw}\right)\right]^{-1} \sum_{d=1}^{D}\sum_{i\in s_d} w_{di}\bX_{di}^t\left(\by_{di}- \bGamma_{dw}(\btheta) \bar{\by}_{dw}\right)^t.
\end{equation}
Using the fact that $\sum_{i\in s_d} w_{di}\bX_{di}=w_{d\cdot}\bar{\bX}_{dw}$, we obtain the alternative expression
\begin{equation}\label{betaMpseudoEBLUP}
    \tilde\bbeta_{w}(\btheta) = \left[\sum_{d=1}^{D}\sum_{i\in s_d} w_{di}\bX_{di}^t \left(\bX_{di}- \bGamma_{dw}(\btheta) \bar{\bX}_{dw}\right)\right]^{-1} \sum_{d=1}^{D}\sum_{i\in s_d} w_{di}\left(\bX_{di}- \bGamma_{dw}(\btheta) \bar{\bX}_{dw}\right)^t \by_{di}.
\end{equation}

Taking expectation in \eqref{betaw} under the MNER model \eqref{MNERdi}--\eqref{MNERdi2} and noting that $E(\by_{di})=\bX_{di}\bbeta$ and $E(\bar\by_{dw})=\bar\bX_{dw}\bbeta$, we obtain that $\tilde\bbeta_w(\btheta)$ is unbiased. Actually, using the alternative expression in \eqref{betaMpseudoEBLUP} and the normality assumptions, we obtain $\tilde\bbeta_w(\btheta)\sim N(\bbeta,\bPhi_w (\btheta))$, with covariance matrix given by
\begin{align}
\bPhi_w (\btheta) &= \left(\sum_{d=1}^D\sum_{i\in s_d}\bX_{di}^t \bA_{di}(\btheta) \right)^{-1}\left\{\sum_{d=1}^D \left[\sum_{i\in s_d}\bA_{di}^t (\btheta)\bSigma_e(\btheta)\bA_{di}(\btheta) \right.\right.\nonumber \\ 
    &+ \left.\left.\sum_{i\in s_d}\bA_{di}^t (\btheta)\bSigma_u(\btheta)\sum_{i\in s_d}\bA_{di}(\btheta)\right]\right\}\left(\sum_{d=1}^D\sum_{i\in s_d} \bA_{di}^t(\btheta)\bX_{di}\right)^{-1},
\end{align}
for $\bA_{di}(\btheta) = w_{di}\left(\bX_{di}-\bGamma_{dw}(\btheta)\bar{\bX}_{dw}\right)$, $i\in s_d$, $d=1\ldots,D$. Replacing $\bbeta$ by $\tilde\bbeta_w(\btheta)$ in \eqref{YRpseudoblup}, we obtain the Multivariate Pseudo-BLUP (MPBLUP) of $\bmu_d=\bar{\bX}_d\bbeta + \bu_d$, denoted $\tilde{\bmu}_d^{MYR} (\btheta)=\tilde{\bmu}_d(\btheta, \tilde{\bbeta}_w(\btheta))$. Similarly, for $\btheta$ we consider a consistent estimator $\hat\btheta$ as $n\to\infty$, such as the REML estimator under the MNER model \eqref{MNERdi}-\eqref{MNERdi2}. Replacing $\btheta$ by $\hat\btheta$ in the MPBLUP, we obtain the Multivariate Pseudo-EBLUP (MPEBLUP) of $\bmu_d$, given by
\begin{equation}\label{YRpseudoEblup}
\hat{\bmu}_d^{MYR} = \tilde{\bmu}_d^{MYR}( \hat{\btheta}) = \hat{\bGamma}_{dw}\big[\bar{\by}_{dw}+(\bar{\bX}_d-\bar{\bX}_{dw})\hat\bbeta_{w}\big]+(\bI_R - \hat{\bGamma}_{dw})\bar{\bX}_d\hat\bbeta_{w},
\end{equation}
where $\hat\bbeta_{w}
= \tilde{\bbeta}_{w}(\hat{\btheta})$ and $\hat{\bGamma}_{dw}= \bGamma_{dw}(\hat\btheta)$.

As in \cite{you2002pseudo}, assume that a constant is included in any of the $R$ model components and that the survey weights satisfy $w_{d\cdot}=N_d$, $d=1,\ldots,D$. In this case, the above Multivariate PEBLUP satisfies the appealing benchmarking property of adding up to the survey regression estimator of the overall total, that is,
\begin{equation*}
\sum_{d=1}^DN_d\hat\bmu_d^{MYR}=\hat\bY+(\bX-\hat\bX)\hat\bbeta_w,
\end{equation*}
where $\bY=\sum_{d=1}^D\sum_{i=1}^{N_d}\by_{di}$ and $\bX=\sum_{d=1}^D\sum_{i=1}^{N_d}\bX_{di}$ are the population totals of the response and auxiliary variables, and $\hat\bY=\sum_{d=1}^D\sum_{i\in s_d}w_{di}\by_{di}$ and $\hat\bX=\sum_{d=1}^D\sum_{i\in s_d}w_{di}\bX_{di}$ are their respective expansion estimators.


Consider now that the survey weights $w_{di}$, $i\in s_d$, were previously calibrated so that $\bar{\bX}_{dw}=\bar{\bX}_{d}$, $d=1,\ldots,D$. They can be calibrated separately for each dimension, obtaining $R$ separate sets of calibrated weights, or jointly for the $R$ dimensions, resulting in a single set of calibrated weights. In both cases, the aggregated model \eqref{FHw} reduces to the MFH model studied by \cite{BENAVENT2016372}, but with general (unknown) covariance matrix for the vector of area effects $\bSigma_u(\btheta)$, and error covariance matrices $\bSigma_{edw}(\btheta)$, $d=1,\ldots,D$, depending on the same parameter vector $\btheta$, for all areas.
%
%
Moreover, the Pseudo BP given in \eqref{YRpseudoblup} reduces to the multivariate BP under the MFH model, given by
\begin{equation}\label{MUpredictor}
\tilde{\bmu}^{MU}_d(\btheta, \bbeta) =\bGamma_{dw}(\btheta)\bar{\by}_{dw}+(\bI_R-\bGamma_{dw}(\btheta))\bar{\bX}_d\bbeta.
\end{equation}
%
This predictor, obtained either from the area-level MFH or the unit-level MNER model \eqref{MNERdi}--\eqref{MNERdi2}, is the multidimensional extension of the ``unified'' predictor of \cite{Acero2025}, and is then called multivariate unified (MU) predictor.
An empirical MU predictor may be obtained fitting the MNER model \eqref{MNERdi}--\eqref{MNERdi2} to the available unit level data, which typically increases the efficiency compared to using area-level data. 

The main advantages of MPEBLUP and the unified predictor over the usual multivariate EBLUP based on a MFH model arise from the specification of the error covariance matrices in terms of a single parameter vector $\btheta$ for all areas in model \eqref{FHw}. This parameterization enables the construction of consistent estimators, as $D \to \infty$, of the error covariance matrices. It also allows us to incorporate the noise of the estimated covariance matrices in the final uncertainty measure of the predictor. This noise is typically ignored in FH and MFH models, but may not be negligible for small $n_d$, as illustrated in \cite{Acero2025}. 

\section{Mean squared error estimation}\label{sec:mse}

Let us first obtain the MSE matrices of the Multivariate Pseudo-Best Predictor $\tilde{\bmu}_d(\btheta,\bbeta)$ and the Multivariate Pseudo-BLUP $\tilde{\bmu}^{MYR}_d(\btheta)$ of $\bmu_d = \bar{\bX}_d\bbeta + \bu_d$.
By \eqref{YRpseudoblup}, we can write $\tilde{\bmu}_d(\btheta,\bbeta) = \bar{\bX}_d \bbeta + \tilde{\bu}_{dw}(\btheta,\bbeta)$, for $\tilde \bu_{dw}(\btheta,\bbeta)=\bGamma_{dw}(\btheta)\left(\bar{\by}_{dw}-\bar{\bX}_{dw}\bbeta\right)$. Then, we have
\begin{eqnarray*}
\tilde{\bmu}_d(\btheta,\bbeta) -\bmu_d=\tilde{\bu}_{dw}(\btheta,\bbeta)-\bu_d=(\bGamma_{dw}(\btheta)-\bI_R)\bu_d+\bGamma_{dw}(\btheta)\bar \be_{dw}.
\end{eqnarray*}
Taking variance above, using \eqref{Vdw} and the expression of $\bGamma_{dw}(\btheta)=\bSigma_{ud}(\btheta)\bV_{dw}^{-1}(\btheta)$, we obtain the MSE matrix of the MPBP $\tilde{\bmu}_d(\btheta,\bbeta)$, given by
\begin{align}
    \mbox{MSE}(\tilde{\bmu}_d(\btheta,\bbeta))&=E\left[\left(\tilde{\bmu}_d(\btheta,\bbeta) - \bmu_d\right)\left(\tilde{\bmu}_d(\btheta,\bbeta) - \bmu_d\right)^t \right] = \var\left(\tilde{\bu}_{dw}(\btheta,\bbeta) - \bu_d\right) \nonumber\\
    &= \left(\bI_R - \bGamma_{dw}(\btheta)\right)\bSigma_{ud}(\btheta) \triangleq \bG_{1d}(\btheta).\label{G1d}
\end{align}

For the MPBLUP $\tilde{\bmu}^{MYR}_d(\btheta)$ , the MSE matrix can be decomposed as
\begin{align}
\mbox{MSE}(\tilde{\bmu}^{MYR}_d(\btheta)) &= \mbox{MSE}(\tilde{\bmu}_d(\btheta,\bbeta))+E\left[\left(\tilde{\bmu}^{MYR}_d(\btheta) - \tilde{\bmu}_d(\btheta,\bbeta)\right)\left(\tilde{\bmu}^{MYR}_d(\btheta) - \tilde{\bmu}_d(\btheta,\bbeta)\right)^t \right] \nonumber \\ &+
E\left[\left(\tilde{\bmu}^{MYR}_d(\btheta) - \tilde{\bmu}_d(\btheta,\bbeta)\right)\left(\tilde{\bmu}_d(\btheta,\bbeta) - \bmu_d\right)^t \right] \nonumber\\ &+
E\left[\left(\tilde{\bmu}_d(\btheta,\bbeta) - \bmu_d\right)\left(\tilde{\bmu}^{MYR}_d(\btheta) - \tilde{\bmu}_d(\btheta,\bbeta)\right)^t \right].\label{decompMSEMYR}
\end{align}
The last two terms in \eqref{decompMSEMYR} are $R\times R$ zero matrices. This can be seen using the law of iterated expectations. Since  $\tilde{\bmu}^{MYR}_d(\btheta)-\tilde{\bmu}_d(\btheta,\bbeta)$ is a function of $\tilde\bbeta_w(\btheta)$ and the data $\by_{w}=(\bar\by_{1w}^t,\ldots,\bar\by_{Dw}^t)^t$, we can write 
\begin{align*}
& E\left[\left(\tilde{\bmu}^{MYR}_d(\btheta) - \tilde{\bmu}_d(\btheta,\bbeta)\right)\left(\tilde{\bmu}_d(\btheta,\bbeta) - \bmu_d\right)^t \right]\\
& \quad  = E_{\by_w,\tilde\bbeta_w(\btheta)}\left\{
\left(\tilde{\bmu}^{MYR}_d(\btheta) - \tilde{\bmu}_d(\btheta,\bbeta)\right)E\left[\left(\tilde{\bmu}_d(\btheta,\bbeta) - \bmu_d\right)^t|\by_w,\tilde\bbeta_w(\btheta)\right]\right\} =\cero_{R\times R},
\end{align*}
since $E(\bmu_d|\by_w,\tilde\bbeta_w(\btheta))=E(\bmu_d|\bar\by_{dw})=\tilde{\bmu}_d(\btheta,\bbeta)$ under model \eqref{FHw}.
Moreover, noting that 
$$
\tilde{\bmu}^{MYR}_d(\btheta) - \tilde{\bmu}_d(\btheta,\bbeta)=\left(\bar{\bX}_d-\Gamma_{dw}(\btheta)\bar{\bX}_{dw}\right)\left(\tilde\bbeta_w(\btheta)-\bbeta\right),
$$
we obtain the second term on the right-hand side of \eqref{decompMSEMYR}, which reads
\begin{align}
& E\left[\left(\tilde{\bmu}^{MYR}_d(\btheta) - \tilde{\bmu}_d(\btheta,\bbeta)\right)\left(\tilde{\bmu}^{MYR}_d(\btheta) - \tilde{\bmu}_d(\btheta,\bbeta)\right)^t \right] \nonumber\\
&\quad\quad = \left(\bar{\bX}_d-\bGamma_{dw}(\btheta)\bar{\bX}_{dw}\right)^t \bPhi_w (\btheta)\left(\bar{\bX}_d-\bGamma_{dw}(\btheta)\bar{\bX}_{dw}\right) \triangleq \bG_{2d}(\btheta).\label{G2d}
\end{align}
From \eqref{G1d} and \eqref{G2d}, we obtain the following expression for the MSE matrix of the MPBLUP $\tilde{\bmu}^{MYR}_d(\btheta)$, 
$$
\mbox{MSE}(\tilde{\bmu}^{MYR}_d(\btheta))=\bG_{1d}(\btheta)+\bG_{2d}(\btheta).
$$

For the MPEBLUP $\hat{\bmu}^{MYR}_d=\tilde{\bmu}^{MYR}_d(\btheta)$, an exact analytical expression for the MSE matrix is not available. In the univariate case,
\cite{you2002pseudo} approximated the MSE along the lines of \cite{prasad1990estimation} and provided an MSE estimator, when using Henderson III estimators of the variance components. For general fitting methods, \cite{Acero2025} proposed a parametric bootstrap MSE estimator. Here we extend the bootstrap procedure of \cite{Acero2025} to the multidimensional case, which is applicable for general fitting procedures.
\vspace{0.3 cm}\\
\noindent {\bf Parametric bootstrap estimator of MSE matrix:}
\begin{enumerate}
\item
Fit the MNER model \eqref{MNERdi}--\eqref{MNERdi2} to the unit-level survey data $\{(\by_{di}, \bX_{di}): i\in s_d, d=1,\ldots,D\}$, obtaining
consistent estimators $\hat{\btheta}$ and $\hat{\bbeta}_w$ of $\btheta$ and $\bbeta$, respectively. 
\item
For $b=1,\ldots,B$, do:
\begin{enumerate}
\item
Taking $\hat{\bSigma}_u=\bSigma_u(\hat\btheta)$ and $\hat{\bSigma}_e=\bSigma_e(\hat\btheta)$, generate bootstrap vectors of area effects and errors for the sampled units, as
$$
\bu_d^{*(b)}\stackiid N_R(\cero_R,\hat{\bSigma}_{u}),\
\be_{di}^{*(b)}\stackiid N_R(\cero_R,\hat{\bSigma}_{e}),\ i\in s_d,\ d=1,\ldots,D.
$$
\item
With the generated area effects $\bu_d^{*(b)}$ from (a) and the fitted model parameters obtained in Step 1, obtain the bootstrap vector of true area means as
$$
\bmu_d^{*(b)}=\bar{\bX}_d\hat\bbeta_w+\bu_d^{*(b)},\ d=1,\ldots,D.
$$
\item
Generate bootstrap vectors of response variables for the sample units as
\begin{equation}\label{bootBNER1}
\by_{di}^{*(b)}=\bX_{di}\hat\bbeta_w+\bu_d^{*(b)}+\be_{di}^{*(b)},\ i\in s_d,\ d=1,\ldots,D.
\end{equation}
\item
Fit the MNER model (\ref{bootBNER1}) to the bootstrapped unit-level data $\{(\by_{di}^{*(b)}, \bX_{di}): i\in s_d, d=1,\ldots,D\}$, obtaining bootstrap parameter estimates $\hat{\btheta}^{*(b)}$ and $\hat{\bbeta}_w^{*(b)}$.
\item
Taking $\bar{\by}_{dw}^{*(b)}=w_{d\cdot}^{-1}\sum_{i\in s_d}w_{di}\by_{di}^{*(b)}$, $\hat{\bSigma}_u^{*(b)}=\bSigma_u(\hat\btheta^{*(b)})$, $\hat{\bSigma}_e^{*(b)}=\bSigma_e(\hat\btheta^{*(b)})$, $\hat\bSigma_{edw}^{*(b)}=k_d^2\hat{\bSigma}_e^{*(b)}$,  $\hat\bV_{dw}^{*(b)}=\hat{\bSigma}_u^{*(b)}+\hat\bSigma_{edw}^{*(b)}$ and $\hat{\bGamma}_{dw}^{*(b)}=\hat{\bSigma}_u^{*(b)}\left(\hat\bV_{dw}^{*(b)}\right)^{-1}$, $d=1,\ldots,D$, obtain the bootstrap MPEBLUP
$$
\hat{\bmu}^{MYR*(b)}_d=\hat{\bGamma}_{dw}^{*(b)}\big[\bar{\by}_{dw}^{*(b)}+(\bar{\bX}_d-\bar{\bX}_{dw})\hat\bbeta_w^{*(b)}\big]+
(\bI_R-\hat{\bGamma}_{dw}^{*(b)})\bar{\bX}_d\hat\bbeta_w^{*(b)}.
$$
\end{enumerate}
\item
A parametric bootstrap estimator of the $R\times R$ matrix $\mbox{MSE}(\hat{\bmu}^{MYR}_d)$  is
\begin{equation}\label{PBMSE}
\mbox{mse}_{PB}(\hat{\bmu}^{MYR}_d)=\frac{1}{B}\sum_{b=1}^B\left(\hat{\bmu}^{MYR*(b)}_d-\bmu_d^{*(b)}\right)\left(\hat{\bmu}^{MYR*(b)}_d-\bmu_d^{*(b)}\right)^t . 
\end{equation}
\end{enumerate}

Using the calibrated weights that satisfy $\bar{\bX}_{dw}=\bar{\bX}_{d}$, $d=1,\ldots,D$, instead of the original weights, in the above bootstrap procedure, we obtain a bootstrap estimator $\mbox{mse}_{PB}(\hat{\bmu}^{MU}_d)$ of the MSE matrix of the multivariate unified predictor, $\hat{\bmu}_d^{MU}$. 

\section{Simulation experiments}\label{sec:sim}

This section describes two simulation experiments, conducted for the following purposes:
\begin{itemize}
\item[A)] To compare the properties (bias and MSE) of alternative survey-weighted estimators of a vector of domain means $\bmu_d$ of $R=2$ response variables. Concretely, on the one hand, we wish to compare the proposed Multivariate Pseudo-EBLUP based on the MNER model, $\hat{\bmu}_{d}^{MYR}$, with the usual EBLUP under an area-level MFH model, denoted $\hat{\bmu}_{d}^{MFH}$. On the other hand, we are also interested in analyzing the gains due to using a multivariate model compared with separate univariate models, so we compare the proposed predictor with the vector of Pseudo-EBLUPs based on independent univariate models for each response variable, denoted $\hat{\bmu}_{d}^{UYR}$. We also compare all these estimators with the weighted direct estimator, $\hat{\bmu}_d^{DIR} = \bar{\by}_{dw}$.
\item[B)] To study the performance of the proposed parametric bootstrap estimator of the MSE matrix of the Multivariate Pseudo-EBLUP, $\hat{\bmu}_d^{MYR}$.
\end{itemize}


%
In the two experiments, we consider populations composed of $D=50$ areas, with population sizes $N_d=500$, $d=1,\ldots,D$, which makes a total of $N=25,000$ population units. 
Vectors of area effects and errors are generated in each Monte Carlo simulation as ${\bu}_{d}\sim N_{2}(\cero,\bSigma_{u}(\btheta))$ and $\be_{di}\sim N_{2}(\cero,\bSigma_{e}(\btheta))$, $i=1,\ldots,N_d$, $d=1,\ldots,D$, where $\btheta = (\sigma^2_{u,1},\sigma^2_{u,2},\sigma_{u,12}, \sigma^2_{e,1},\sigma^2_{e,2},\sigma_{e,12})^t=(0\mbox{.}1,0\mbox{.}4,0\mbox{.}16,0\mbox{.}9,1,0\mbox{.}75)^t$.

We consider a single covariate and a constant for each dimension, that is, $p_1=p_2=2$, with vectors of regression coefficients for each dimension given by
$\bbeta_1=(\beta_{11},\beta_{12})^t =(1,1)^t$ and
$\bbeta_2=(\beta_{21},\beta_{22})^t =(4,0.5)^t$.
We take vectors of covariates $\bx_{di1}=(x_{di11},x_{di12})^t$ and $\bx_{di2}=(x_{di21},x_{di22})^t$ with $x_{di11}=x_{di21}=1$, and generate $x_{di12}\sim \mbox{Gamma}(k=2, \alpha=5)$ and $x_{di22}\sim \mbox{Gamma}(k=5+3d/D,\alpha=5)$, $i=1,\ldots,N_d$, $d=1,\ldots,D$, where here $k$ and $\alpha$ denote the shape and scale parameters, respectively. Then we take $\bX_{di}=\mbox{diag}(\bx_{di1}^t, \bx_{di2}^t)_{2\times 4}$, $i=1,\ldots,N_d$, $d=1,\ldots,D$.


The sample $s$ is drawn by simple random sampling without replacement, independently from each area $d$, with area sample sizes $n_d\in \{5,10,15,20,25\}$, where each different value is repeated for 10 consecutive areas. The sampling weights are then $w_{di}=N_d/n_d$, $i \in s_d$, $d=1,\ldots,D$. We keep the covariate values $\bX_{di}$ for all $i$ and $d$ and the sample $s=s_1\cup\ldots\cup s_D$ fixed throughout the simulation replicates.

Using the above setup, in each Monte Carlo simulation $\ell=1,\ldots,L$, for $L=1,000$, we generate area effects $\bu_d^{(\ell)}\sim N_2(\cero,\bSigma_u(\btheta))$ and errors $\be_{di}^{(\ell)}\sim N_{2}(\cero,\bSigma_e(\btheta))$, for all population units $i=1,\ldots,N_d$, $d=1,\ldots,D$. From them, we obtain the population of response vectors as $\by_{di}^{(\ell)}=\bX_{di} \bbeta+\bu_{d}^{(\ell)}+\be_{di}^{(\ell)}$, $i=1,\ldots,N_d$, $d=1,\ldots,D$. From these values, we compute the true mean vectors, 
$$
\bmu_{d}^{(\ell)}= \frac{1}{N_d}\sum_{i=1}^{N_d}\by_{di}^{(\ell)},\quad d=1,\ldots,D.
$$
Then, taking the values $\by_{di}^{(\ell)}$ for the sample units $i \in s_d$, $d=1,\ldots,D$, we compute the corresponding estimators $\hat{\bmu}_d^{DIR(\ell)}$, $\hat{\bmu}_d^{MFH(\ell)}$, $\hat{\bmu}_d^{MYR(\ell)}$, and $\hat{\bmu}_d^{UYR(\ell)}$, $d=1,\ldots,D$. The model-based estimators were obtained using the REML fitting method.

For a generic estimator $\hat{\mu}_{dr}$ of the (univariate) domain mean $\mu_{dr}$, $r=1,2$, we evaluate its performance in terms of relative bias (RB) and relative root mean squared error (RRMSE), approximated empirically as
\begin{equation*}
	\mbox{RB}(\hat{\mu}_{dr}) =  \frac{ L^{-1} \sum_{\ell=1}^{L} (\hat{\mu}_{dr}^{(\ell)} - \mu_{dr}^{(\ell)})}{L^{-1} \sum_{\ell=1}^{L} \mu_{dr}^{(\ell)}}, \quad \mbox{RRMSE}(\hat{\mu}_{dr}) = \frac{ \sqrt{ L^{-1} \sum_{\ell=1}^{L} (\hat{\mu}_{dr}^{(\ell)} - \mu_{dr}^{(\ell)})^2} }{L^{-1} \sum_{\ell=1}^{L} \mu_{dr}^{(\ell)}}.
\end{equation*}
We also summarize these measures by taking averages of absolute relative bias, $|\mbox{RB}(\hat{\mu}_{dr})|$, and $\mbox{RRMSE}(\hat{\mu}_{dr})$, over areas $d$ with the same sample size, for $r=1,2$.

\begin{figure}[H]%
	\centering	\includegraphics[width=150mm]{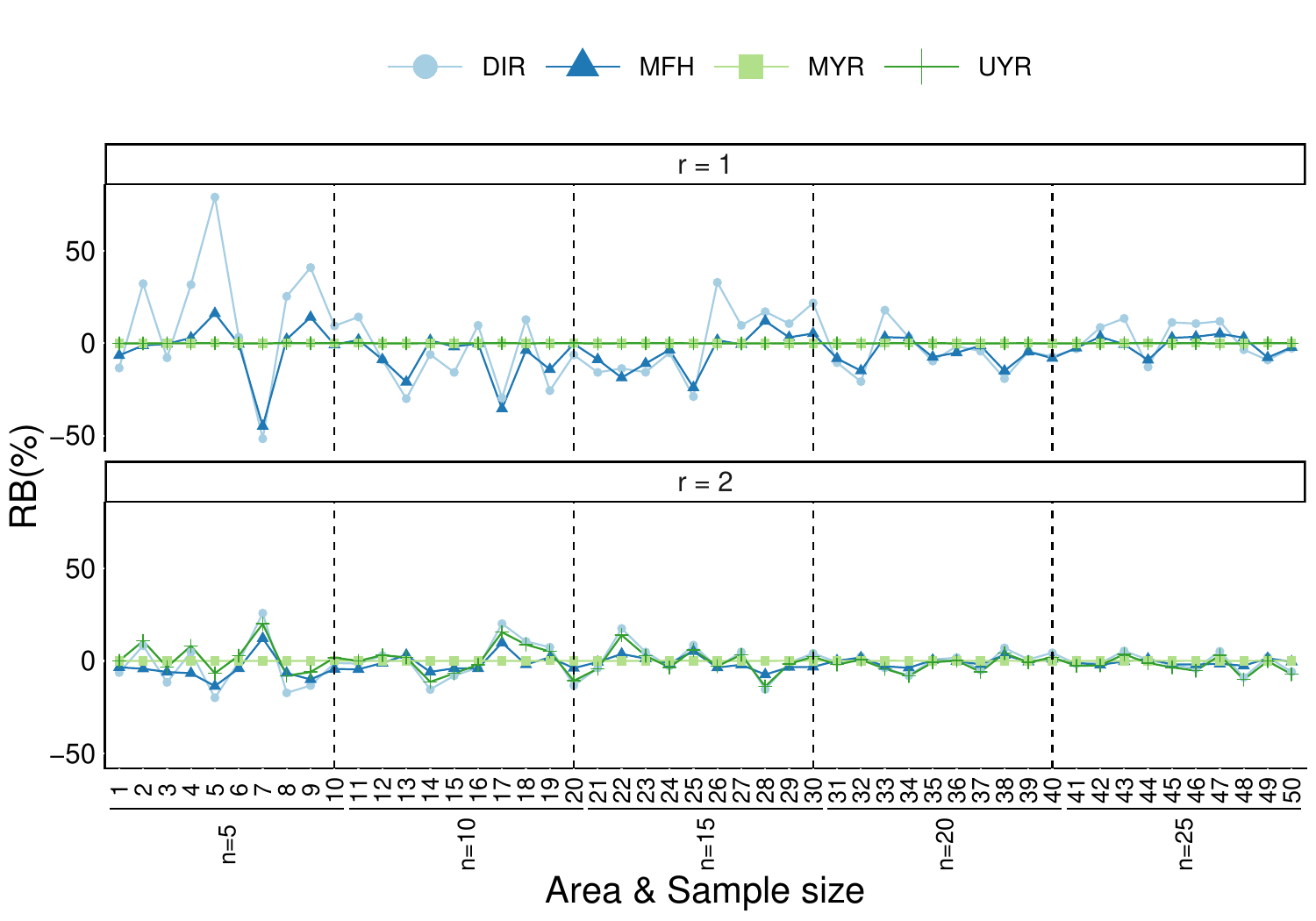}
	\caption{\% RB of DIR, MFH, MYR and UYR, for $d=1,\ldots, D$, for $r=1$ (above) and $r=2$ (below).}\label{fig:RB_50}
\end{figure}

\begin{figure}[H]%
	\centering	\includegraphics[width=150mm]{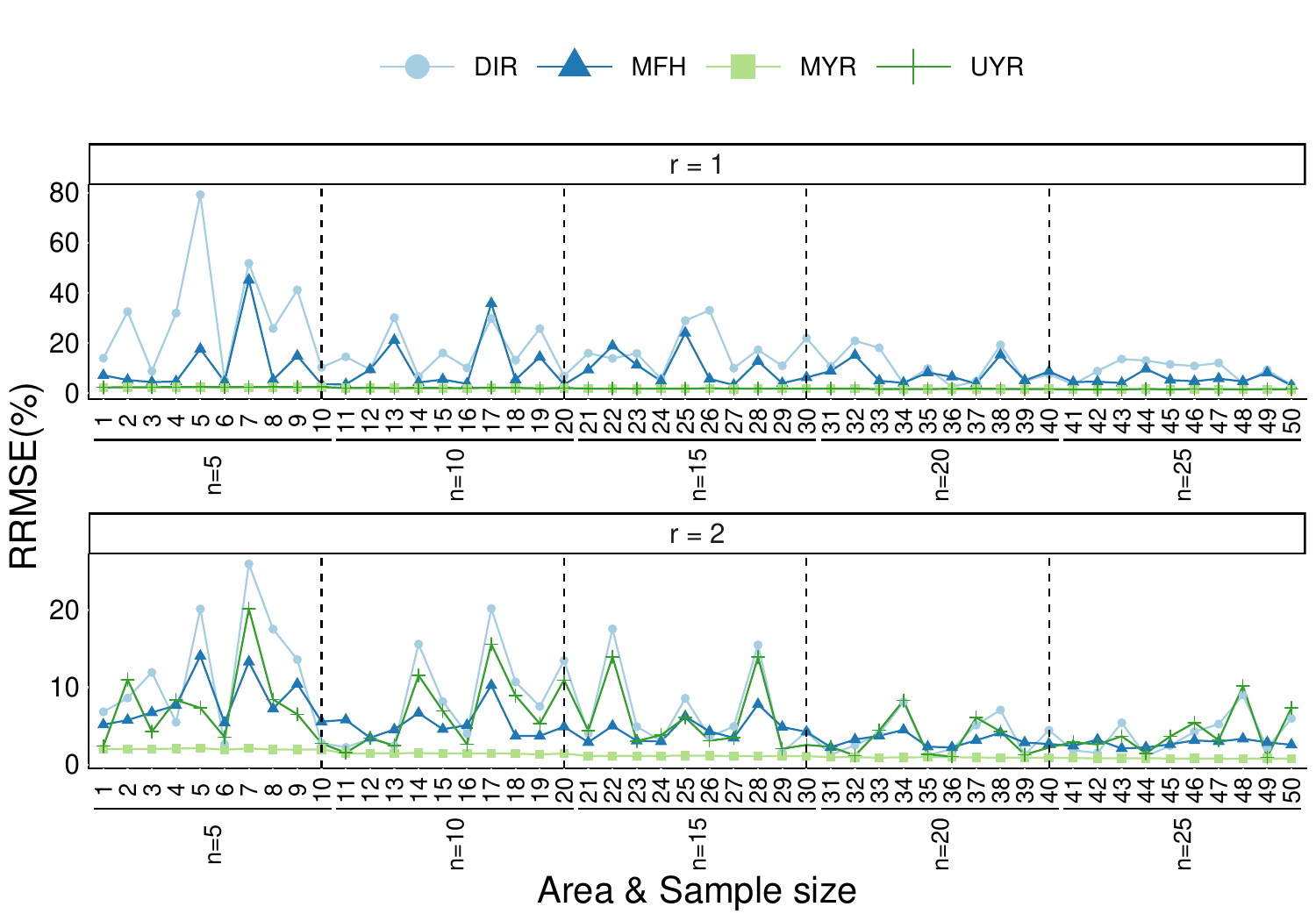}
	\caption{\% RRMSE of DIR, MFH, MYR and UYR for each area $d=1,\ldots,D$, for $r=1$ (above) and $r=2$ (below).}\label{fig:RRMSE_50}
\end{figure}

Figures \ref{fig:RB_50} and \ref{fig:RRMSE_50} display respectively the resulting values of RB (\%) and RRMSE (\%) of each estimator in the $y$ axis, for each area $d=1,\ldots,D$, in the $x$-axis. These figures 
show the typical instability of the direct estimators (or large RRMSE) for areas with small sample sizes, compared to the other estimators. They also illustrate the efficiency gains of the usual EBLUP under the MFH model with respect to the direct estimator, for practically all areas. The univariate Pseudo-EBLUP performs well for the first component ($r=1$), but not for the second component, for which the predictive power of the model is weaker, because $\sigma_{u,2}^2>\sigma_{u,1}^2$ in this simulation experiment.
Finally, the Multivariate Pseudo-EBLUP, $\hat{\bmu}_d^{MYR}$, performs the best in terms of ARB and RRMSE for all areas and for the two components of $\bmu_d$. This is somewhat expected, since this predictor exploits the unit-level data of much larger size ($n=750>D=50$), which increases the efficiency compared to the EBLUP under the area-level MFH model. It also accounts for the correlation between response variables, which increases the efficiency of the predictor of the mean for the second component by borrowing strength from the first component, yielding an improvement over the Pseudo-EBLUP based on the univariate model for that component. See how, in this second component, the univariate Pseudo-EBLUP actually tends to perform worse than the EBLUP under the MFH area-level model, because the latter also exploits the correlation among the two response components.


Similar conclusions are drawn looking at the averages of the absolute relative bias ($\overline{\mbox{ARB}}$), and of the RRMSE ($\overline{\mbox{RRMSE}}$) over areas with the same sample size, reported in Table \ref{tab:tab_RB_RRMSE}, for the four estimators of $\mu_{d1}$ ($r=1)$ and $\mu_{d2}$ ($r=2)$. The gains of the Multivariate Pseudo-EBLUP over the direct estimator and the EBLUP under the MFH model are striking, as well as the gains over the univariate Pseudo-EBLUP for $\mu_{d2}$.

\begin{longtable}{ccccccc ccccc}
    \caption{Average across areas with the same sample size of \% ARB and RRMSE for DIR, MFH, MYR and UYR.}
    \label{tab:tab_RB_RRMSE} \\
    
    \toprule
    & & \multicolumn{5}{c}{$\overline{\text{ARB}}$} & \multicolumn{5}{c}{$\overline{\text{RRMSE}}$} \\
    \cmidrule(lr){3-7} \cmidrule(lr){8-12}
    Estimator & $r/n_d$ & $5$ & $10$ & $15$ & $20$ & $25$ & $5$ & $10$ & $15$ & $20$ & $25$ \\
    \midrule
    \endfirsthead

    \toprule
    & & \multicolumn{5}{c}{$\overline{\text{ARB}}$} & \multicolumn{5}{c}{$\overline{\text{RRMSE}}$} \\
    \cmidrule(lr){3-7} \cmidrule(lr){8-12}
    Estimator & $r/n_d$ & $5$ & $10$ & $15$ & $20$ & $25$ & $5$ & $10$ & $15$ & $20$ & $25$ \\
    \midrule
    \endhead

    \bottomrule
    \endfoot

    $\hat{\bmu}_d^{DIR}$ & 1 & 29.49 & 15.89 & 17.12 & 9.85 & 8.69 & 30.02 & 16.19 & 17.3 & 10.15 & 8.91 \\
    & 2 & 10.92 & 8.46 & 6.60 & 3.44 & 3.73 & 11.54 & 8.76 & 6.80 & 3.69 & 3.89 \\
    $\hat{\bmu}_d^{MFH}$& 1 & 8.96 & 8.87 & 8.86 & 7.07 & 4.01 & 11.18 & 10.59 & 9.97 & 7.99 & 5.37 \\
    & 2 & 7.15 & 4.06 & 3.19 & 1.58 & 1.45 & 8.13 & 5.26 & 4.48 & 3.10 & 2.74 \\
    $\hat{\bmu}_d^{MYR}$ & 1 & 0.04 & 0.04 & 0.03 & 0.05 & 0.05 & 2.23 & 1.89 & 1.7 & 1.58 & 1.46 \\
    & 2 & 0.04 & 0.04 & 0.04 & 0.02 & 0.02 & 2.00 & 1.43 & 1.11 & 0.92 & 0.80 \\
    $\hat{\bmu}_d^{UYR}$ & 1 & 0.06 & 0.05 & 0.04 & 0.04 & 0.05 & 2.35 & 2.01 & 1.76 & 1.63 & 1.50 \\
    & 2 & 6.76 & 6.54 & 5.43 & 2.90 & 3.93 & 7.47 & 6.92 & 5.63 & 3.23 & 4.12 \\
\end{longtable}

Next, we describe the results of our simulation experiment B, conducted to analyze the performance of the parametric bootstrap estimator of the MSE matrix, $\hbox{mse}_{PB}(\hat{\bmu}^{MYR}_d)$, given in \eqref{PBMSE}. Concretely, we analyze the diagonal elements of this matrix, which estimate the corresponding MSEs of each component of $\hat{\bmu}^{MYR}_d=(\hat{\mu}^{MYR}_{d1},\hat{\mu}^{MYR}_{d2})^t$.

The true MSEs of each component of the Multivariate Pseudo-EBLUP were initially approximated as in the simulation experiment A. Subsequently, a second simulation study was conducted with $L=500$ replicates. For each replicate, we computed the parametric bootstrap MSE estimator $\hbox{mse}_{PB}(\hat{\bmu}^{MYR}_d)$, based on $B=500$ bootstrap samples. 

\begin{figure}[H]%
	\centering
	\includegraphics[width=160mm]{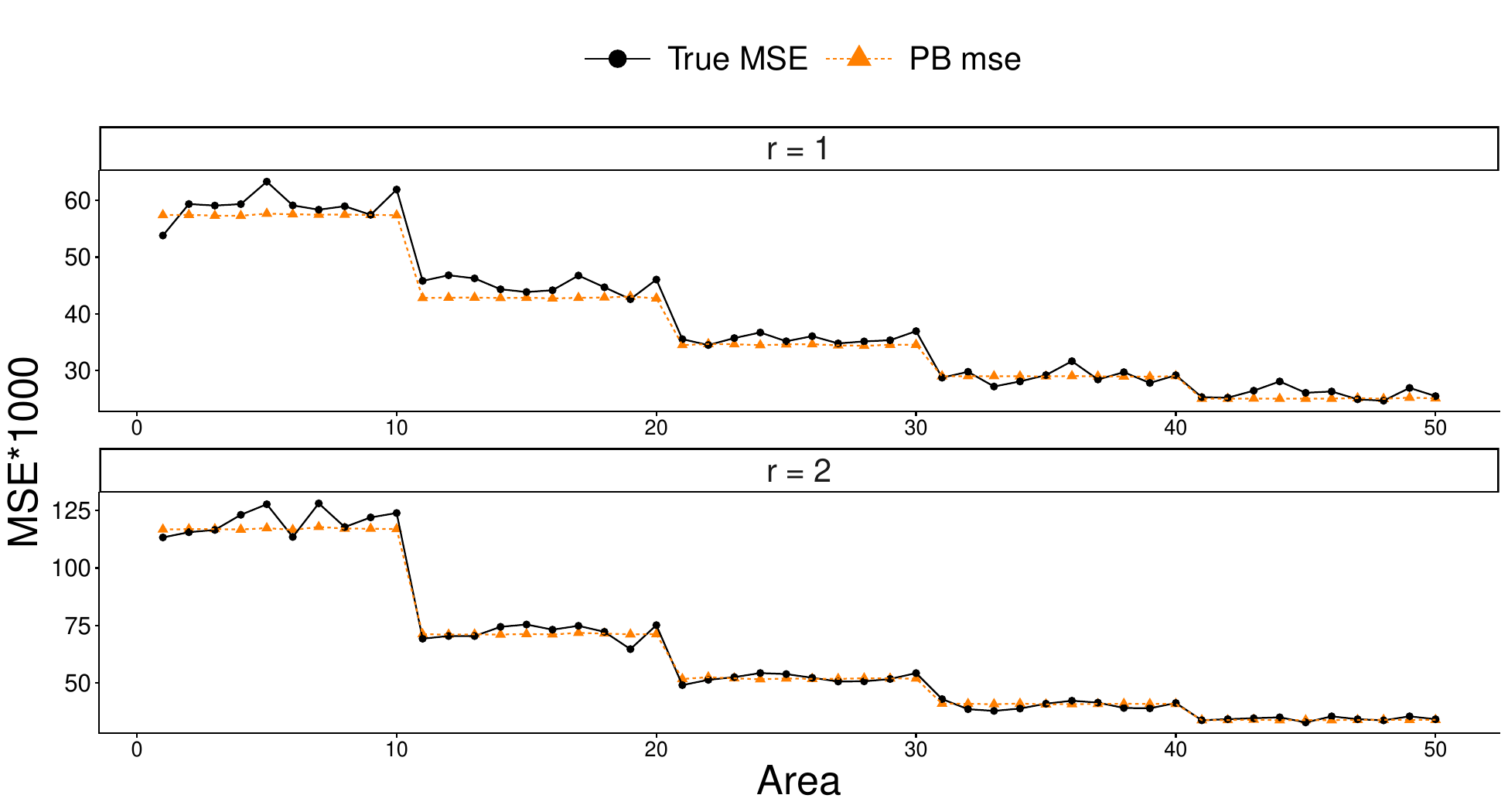}
	\caption{True values of $\mbox{MSE}(\hat{\mu}^{MYR}_{dr})$ and MC averages of PB MSE estimator, $\mbox{mse}_{PB}(\hat{\mu}^{MYR}_{dr})$, for each area $d=1,\ldots,D$, for $r=1$ (above) and $r=2$ (below).}\label{fig:mseplot_MU}
\end{figure}

Figure \ref{fig:mseplot_MU} depicts the true MSE and the Monte Carlo average of the PB MSE estimator ($y$ axis), for each area ($x$ axis), for $\mu_{d1}$ (above) and $\mu_{d2}$ (below). We can see how the PB estimator tracks the true MSEs rather well, at least for areas with not extremely small sample sizes. Hence, the proposed parametric bootstrap method provides a reasonable approach for the estimation of the MSE of each component of the Multivariate Pseudo-EBLUP, applicable with different model fitting methods.

\section{Application: Housing price and rent in Colombia}
\label{application}

The 2023 Survey on Living Conditions (in Spanish, \textit{Encuesta de Calidad de Vida}, ECV), is a national household survey conducted annually by the National Administrative Department of Statistics (DANE in Spanish). This survey quantifies and characterizes the living conditions of households in Colombia and includes variables related to housing (wall and floor materials, access to public, private, and communal services, and social stratification), individuals (education, health, childcare, and the use of information and communication technologies), and households (ownership of property, goods, and perceptions about living conditions within the dwelling, among others). 

The survey sample is drawn by two-stage sampling. In the first stage, municipalities are drawn by probability proportional to size (PPS) sampling. In the second stage, city blocks (composed of 10 households) are drawn according to systematic sampling. 

The ECV collects the monthly rent that homeowners would be willing to pay for their property if they had to, $y_{di1}$, and their actual monthly mortgage payment, $y_{di2}$. 
We wish to estimate the (figurative) monthly rental cost (MRC) of the properties owned, $\mu_{d1}=N_d^{-1}\sum_{i=1}^{N_d}y_{di1}$, and the monthly mortgage payment (MP), $\mu_{d2}=N_d^{-1}\sum_{i=1}^{N_d}y_{di2}$, for the areas $d=1,\ldots,D$, described below. A joint model for monthly rental price and mortgage installment captures housing’s use and ownership values, enabling assessment of whether market valuations in rental and mortgage markets are coherent.

We first define the population of interest. Since we wish to reflect full real market prizes, we select only the property owners who are still paying for their properties.  
Moreover, we restrict the analysis to homeowners who pay quantities not exceeding one million COP in mortgage payments, and who would be willing to pay up to the same amount in rent if they had to, excluding those belonging to the highest socio-economic stratum (stratum 6). One million COP corresponds approximately to one month of minimum wage in 2023. This restriction ensures that the estimated indicators reflect the dynamics of households operating under realistic affordability constraints, excluding luxury or high-end properties that follow different pricing mechanisms. By excluding high-value properties and high-income households, the study concentrates on the segment that is most sensitive to housing affordability and financial stress, ensuring that the derived indicators and policy implications are relevant to middle- and lower-income groups.

The target areas are the crossings of the 32 Colombian departments and the Capital District with household type (H = House, A = Apartment). A total of $D=54$ areas were considered, after excluding high-value properties, high-income households, and cases that were not consistent with the 2018 National Population and Housing Census (CNPV in Spanish), also conducted by DANE, from which we obtained the population means of the auxiliary variables for each area. 

The number of households with homeowners in the ECV is $n=905$. The average area sample size is 17, with a minimum of 2, quartiles $Q_1 = 6$, $Q_2 = 13$ and $Q_3 = 22.5$, and a maximum sample size of 67. The small sample sizes in most areas justifies the application of SAE techniques.

The sample correlation of $y_{di1}$ and $y_{di2}$ is 0.46, indicating a clear positive correlation. Accordingly, we consider a bivariate NER model for $(y_{di1},y_{di1})^t$. In each model component $r=1,2$, we include an intercept and the socio-economic stratum of the household, which has 5 categories after excluding stratum 6. Considering the first stratum as the reference category, we include 4 dummy indicators for the remaining 4 strata, that is, here we have $p_1=p_2=5$, which gives $p=10$. 

\begin{figure}[H]%
	\centering
	\includegraphics[width=160mm]{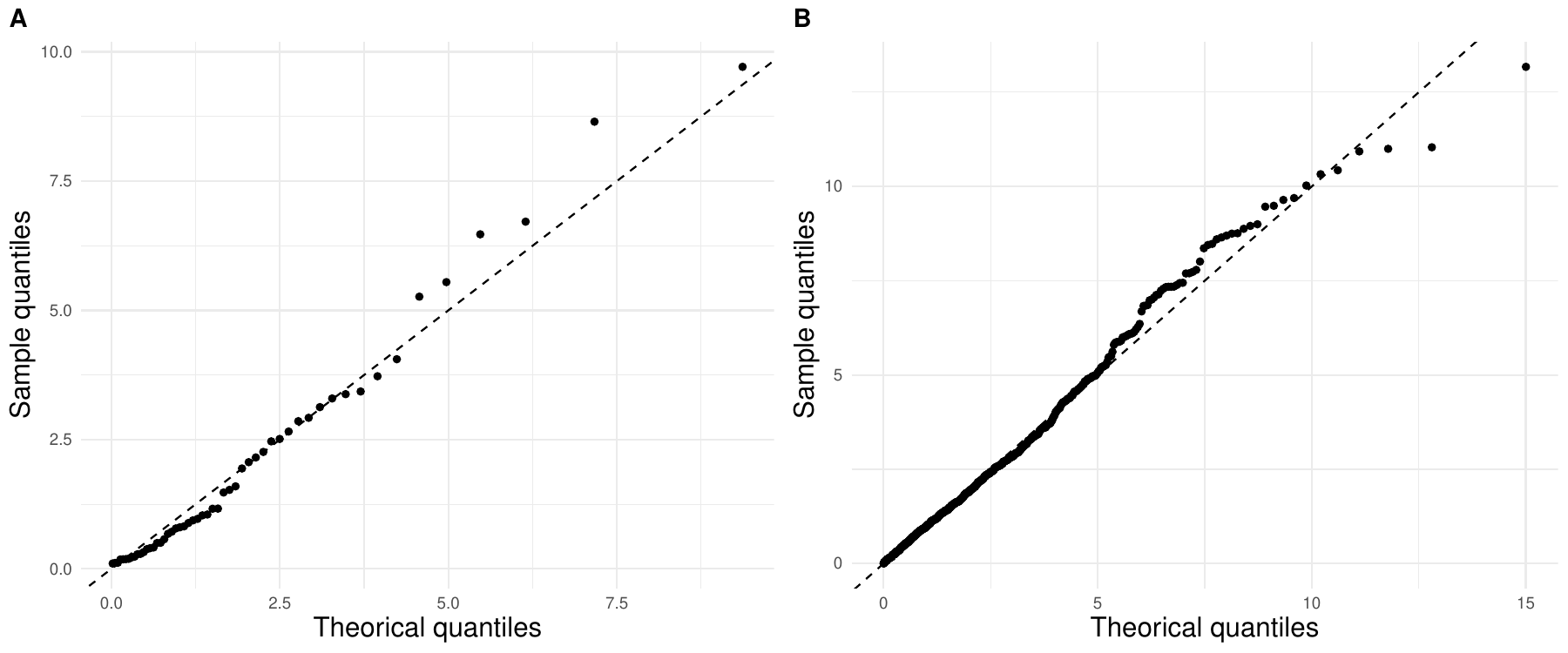}
	\caption{Bivariate Normal Q-Q plot of predicted area effects (A) and Bivariate Normal Q-Q plot of unit-level residuals from BNER model (B).}\label{fig:qqRandRes}
\end{figure}

We fit the BNER model \eqref{mod2NERd} to the ECV survey data using the REML method. For this model, Figure \ref{fig:qqRandRes} shows (A) a bivariate normal Q-Q plot of $\hat \bu_{dw}=\tilde \bu_{dw}(\hat\bbeta_w,\hat\btheta)$, $d=1,\ldots,D$, and (B) a bivariate normal Q-Q plot of unit-level residuals $\hat \be_{di}=\by_{di}-\bX_{di}\hat\bbeta_w-\hat \bu_{dw}$, $i\in s_d$, $d=1,\ldots,D$, both based on the Mahalanobis distance obtained with respective covariance matrices $\hat\bSigma_u$ and $\hat\bSigma_e$. These plots do not indicate serious departures from the bivariate normal assumptions. 

The fitted regression coefficients and variance components are reported respectively in Tables \ref{tab:estimated_Coef_BNER} and \ref{tab:estimated_theta_BNER} of Appendix \ref{sec:plotDiagnosticsUnivariate}. We also fitted univariate NER models for each response variable, to obtain the univariate Pseudo-EBLUPs proposed by \citet{you2002pseudo}, $\hat{\bmu}^{UYR}_d$, which we compare with the proposed Bivariate Pseudo-EBLUP. In these univariate models, we neither found serious departures from the normality assumptions in the area- or unit-level residuals. Diagnostic plots can be seen in Figures \ref{fig:residuals_model_univariate_MRC} and \ref{fig:residuals_model_univariate_MP} of Appendix \ref{sec:plotDiagnosticsUnivariate}, respectively.

\begin{figure}[H]%
	\centering
	\includegraphics[width=160mm]{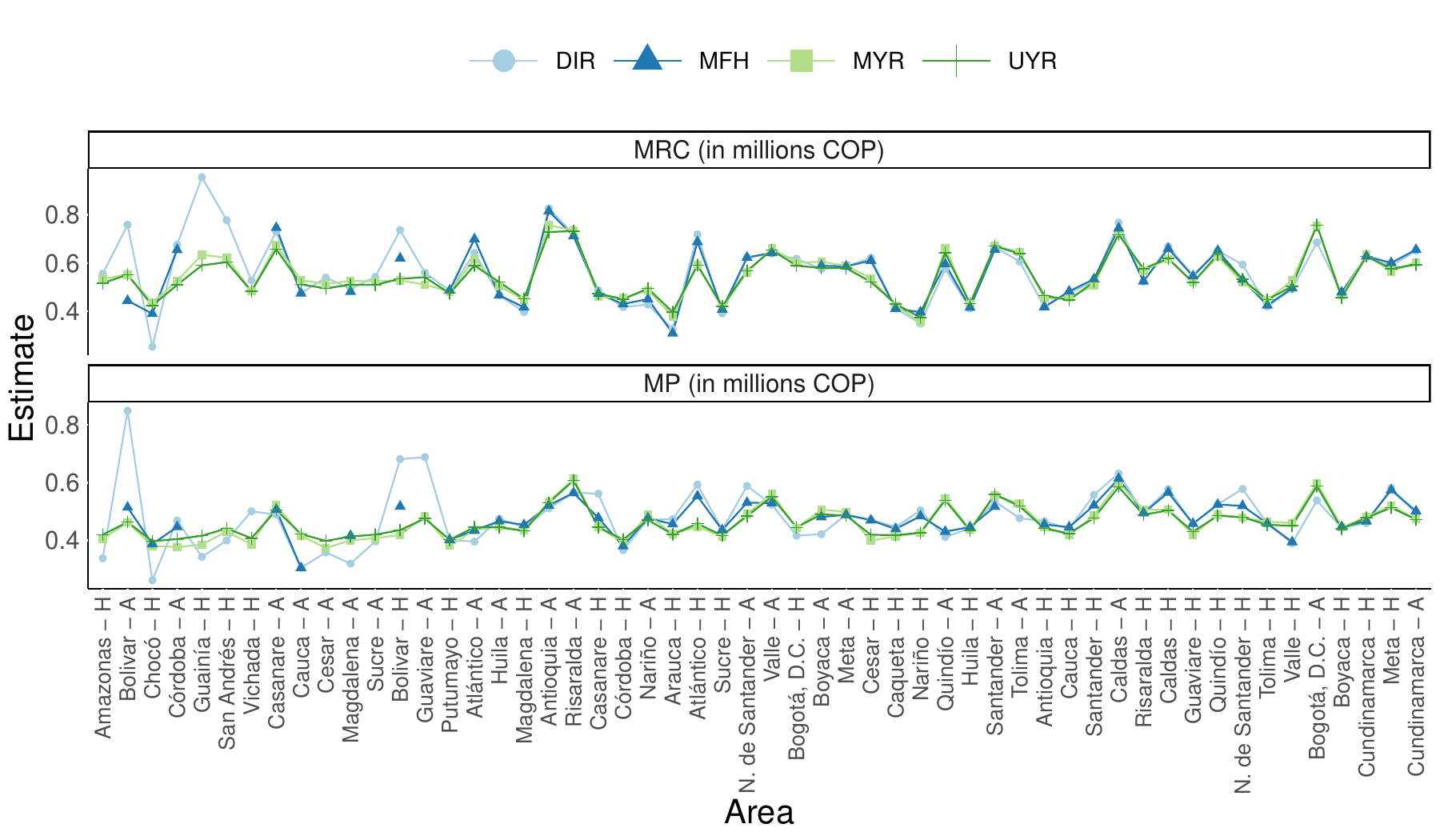}
	\caption{DIR, MFH, MYR and UYR estimators of the means of MRC and MP by department and household type, presented in ascending order of sample size.}\label{fig:estimates}
\end{figure}

Figure \ref{fig:estimates} depicts, above and below respectively, the resulting values for the two components of the four different predictors of $\bmu_d$ considered in our simulation experiments, DIR, MFH, MYR and UYR, with areas sorted in the ascending order of sample size. 
As can be seen in the figure, for some areas with very small sample sizes, the estimated variances of the direct estimators turn out to be very almost equal to zero, which is completely unreasonable. These areas were not included in the MFH model, and hence the values of the estimator $\hat{\boldsymbol{\mu}}_{d}^{MFH}$ are not plotted for them.

Observe again the instability of the direct estimator for areas with very small sample sizes such as in Bolívar--Apartment. All other estimators show a more stable behavior, with MYR and UYR the most stable ones and taking here pretty similar values.

\begin{figure}[H]%
	\centering
	\includegraphics[width=160mm]{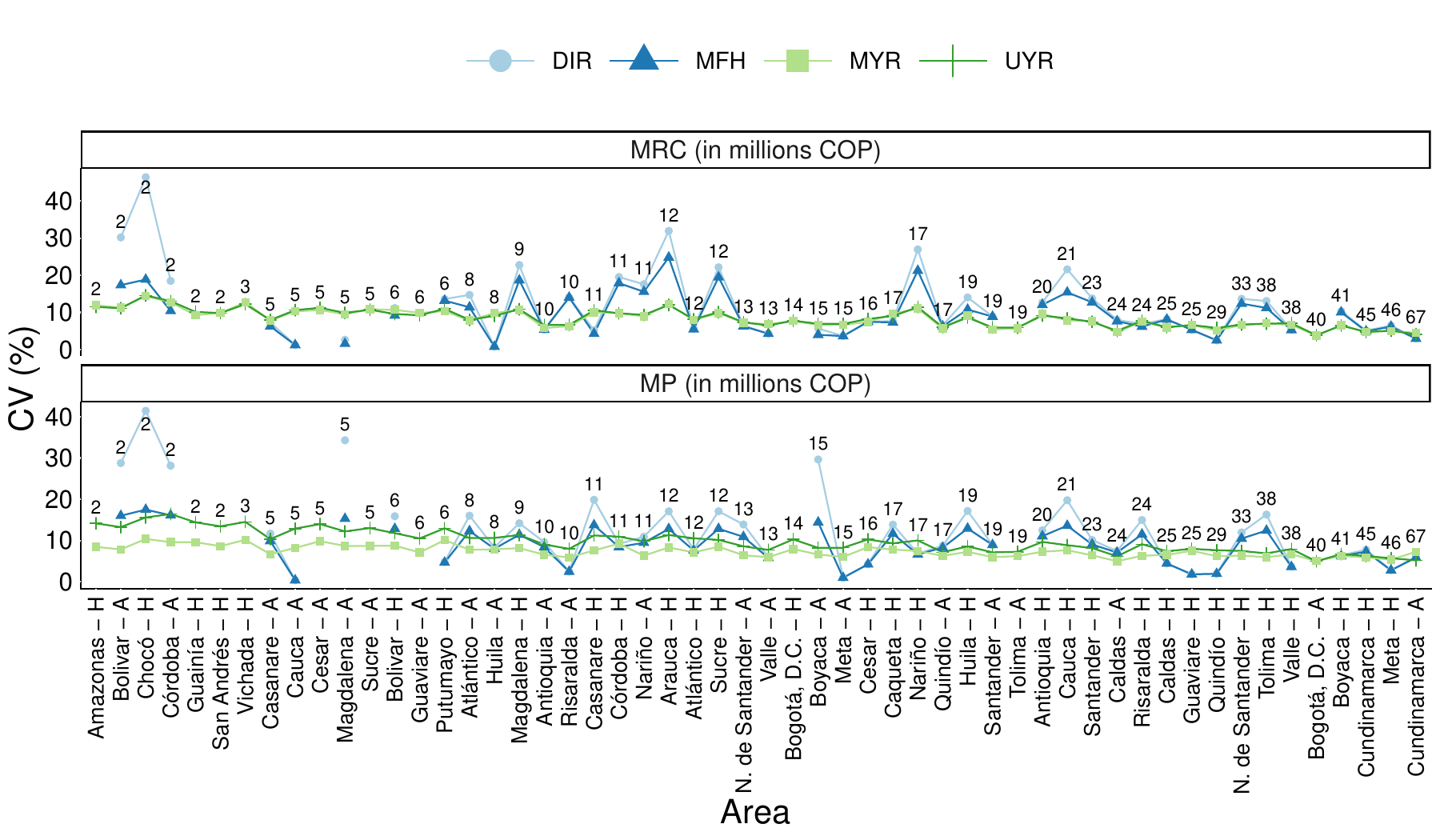}
	\caption{Estimated CVs of DIR, MFH, MYR and UYR estimators of the Monthly Rental Cost (MRC) and Mortgage Payment (MP), with areas sorted in the ascending sample size $n_d$ (in point labels).}\label{fig:mse_estimates}
\end{figure}

Figure \ref{fig:mse_estimates} displays the estimated CV of each estimator, for each area. For the direct estimator, it is the squared root of the estimated design-variance by the estimator. 
For the model-based estimators, the estimated CV is obtained as the squared root of the estimated model MSE, by the corresponding estimator. In the case of MFH, the estimated MSE is obtained by the analytical approximation obtained by \cite{BENAVENT2016372}. For MYR, the MSE was estimated using the bootstrap method introduced in Section \ref{sec:mse}. Finally, for UYR, we used the proposal of \cite{Acero2025}, which is a special case with $R=1$ of our proposed bootstrap method. 

Again, for the areas where the estimated variance of the direct estimator is nearly zero, estimated CVs are not  displayed for DIR and MFH in Figure \ref{fig:mse_estimates}. In these areas, MYR and UYR actually obtain more reasonable smoother CV estimates. For the MP response variable (shown below), we can see that the estimated CVs obtained from the bivariate NER model (MYR estimator) are lower than the estimated CVs obtained from independent NER models (UYR estimator). As illustrated in our simulation experiment, this occurs because the univariate model is weaker for the MP variable. Note from Table \ref{tab:estimated_theta_BNER} in Appendix \ref{sec:plotDiagnosticsUnivariate} that $\hat{\sigma}_{e,2}^2 > \hat{\sigma}_{e,1}^2$. The BNER model borrows strength from the other response variable, MRC, thereby increasing the efficiency of the MYR predictors of the MP mean. 

\section{Conclusions and discussion}\label{sec:conclusions}

This paper extends the Pseudo-EBLUP of \cite{you2002pseudo} to the case of multiple correlated characteristics, which ``borrows strength'' also from the different characteristics. The proposed predictor is based on the assumption of a multivariate NER model, which is aggregated to the area level using the survey weights. 

If the survey weights had been calibrated for each area, so that $\bar{\bX}_{dw}=\bar{\bX}_{d}$, $d=1,\ldots, D$, the survey aggregated model becomes a type of MFH model, whose error covariance matrices depend on common model parameters for all areas. In this case, the proposed predictor becomes a multidimensional extension of the Unified predictor of \cite{Acero2025}.


We propose a parametric bootstrap method to estimate the MSE matrix of the Multivariate Pseudo-EBLUP and the Unified predictor, which is applicable for general fitting procedures. 

The results from our simulation experiments and our application yield similar conclusions, indicating gains over the univariate Pseudo-EBLUPs when the corresponding univariate model has weaker predictive power, provided that the response variables are correlated. They also indicate gains over the MFH model, due to the use of larger sample size, and more informative unit-level survey data, to fit the MNER model.


\section*{Appendix}
\appendix

\section{Additional application results}\label{sec:plotDiagnosticsUnivariate}

\begin{figure}[H]%
	\centering
	\includegraphics[width=150mm]{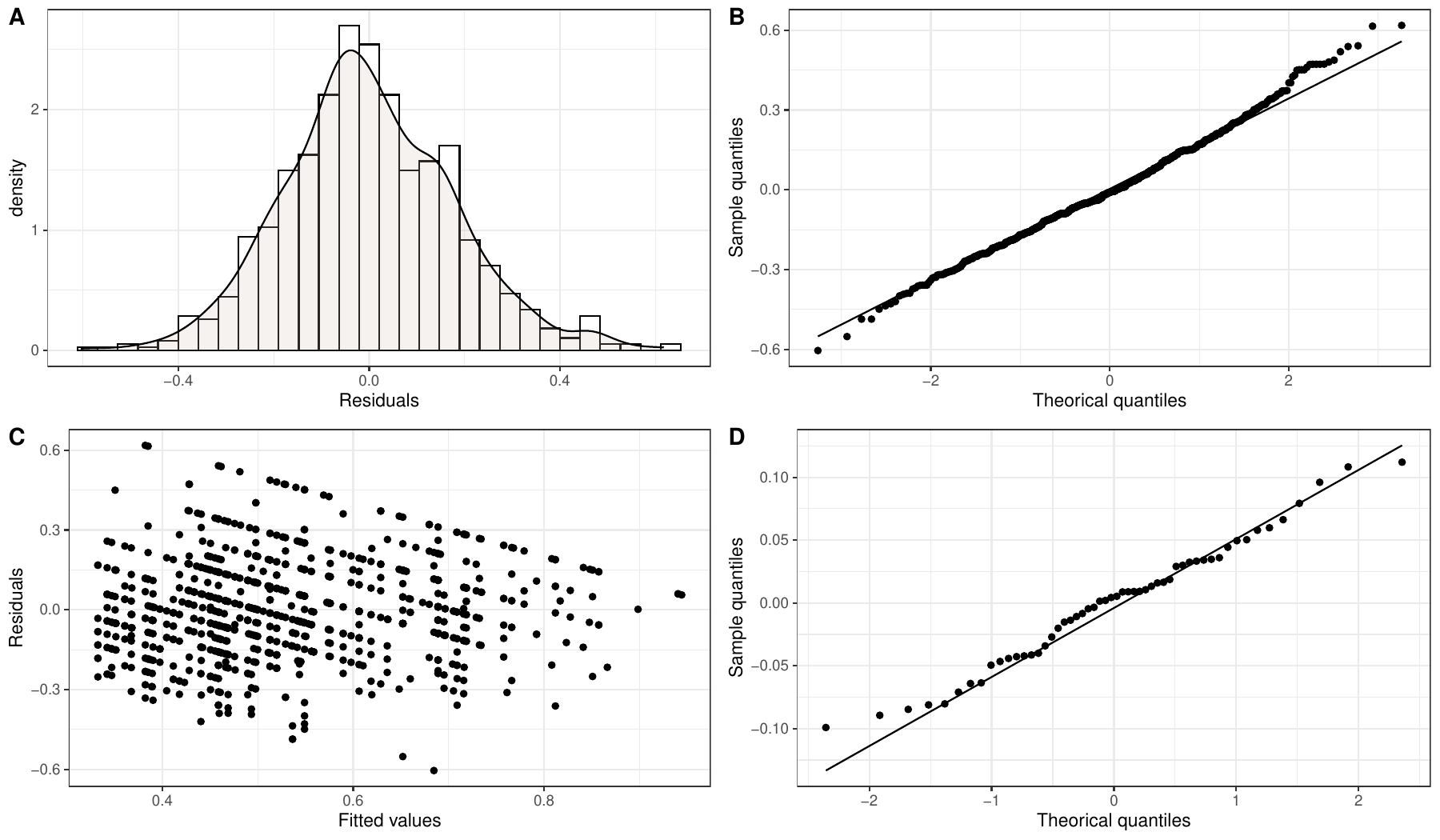}
	\caption{Histogram of unit-level residuals (A), Normal Q-Q plot of unit-level residuals (B), scatterplot of unit-level residuals against predicted values (C) and Normal Q-Q plot of predicted area effects (D) in the NER model for MRC.}\label{fig:residuals_model_univariate_MRC}
\end{figure}

\begin{figure}[H]%
	\centering
	\includegraphics[width=150mm]{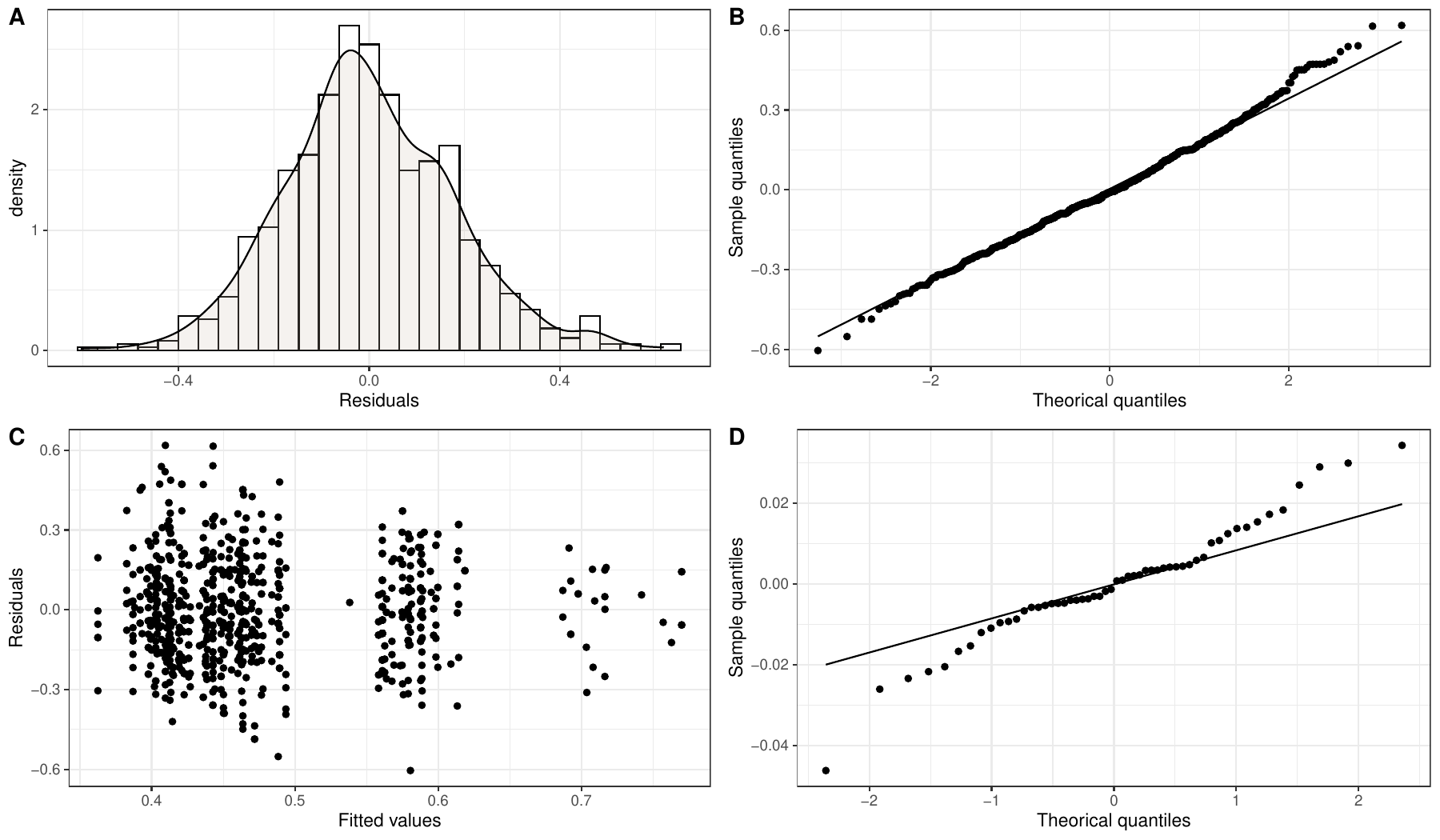}
	\caption{Histogram of unit-level residuals (A), Normal Q-Q plot of unit-level residuals from NER model (B), scatterplot of unit-level residuals against predicted values (C) and Normal Q-Q plot of predicted area effects (D) from the NER model for MP.}\label{fig:residuals_model_univariate_MP}
\end{figure}

\begin{table}[H]
\centering
\caption{Fitted regression coefficients $\hat{\bbeta}_w$ for the BNER model for MRC and MP.}
\label{tab:estimated_Coef_BNER}
\begin{tabular}{cccccc}
\hline
Response             & Covariate  & Estimate & Std.error & t.value & p.value \\ \hline
\multirow{5}{*}{MRC} & (Intercept) & 0.48 & 0.02 & 22.12 & 0.000 \\
                     & Strata 2    & 0.11 & 0.02 & 4.64 & 0.000 \\
                     & Strata 3    & 0.26 & 0.03 & 8.81 & 0.000 \\
                     & Strata 4    & 0.38 & 0.05 & 7.19 & 0.000 \\
                     & Strata 5    & 0.33 & 0.09 & 3.80 & 0.000 \\
                     \hline
\multirow{5}{*}{MP}  & (Intercept) & 0.38 & 0.02 & 16.78 & 0.000   \\
                     & Strata 2    & 0.10 & 0.03 & 3.49 & 0.001   \\
                     & Strata 3    & 0.27 & 0.03 & 7.59 & 0.000   \\
                     & Strata 4    & 0.36 & 0.06 & 5.73 & 0.000   \\
                     & Strata 5    & 0.42 & 0.10 & 4.03 & 0.000   \\ \hline
\end{tabular}
\end{table}

\begin{table}[H]
\centering
\caption{Fitted parameters $\hat{\btheta}$ of the BNER model for MRC and MP.}
\label{tab:estimated_theta_BNER}
\begin{tabular}{ccccc}
\hline
Parameter & Estimate & Std.error & t.value  & p.value \\ \hline
$\sigma_{u,1}^2$  & 0.0051  & 0.0000 & 2245.5748  & 0.0000 \\
$\sigma_{u,2}^2$  & 0.0011  & 0.0000 & 1887.2100  & 0.0000 \\
$\sigma_{u,12}$   & -0.0004 & 0.0000 & -641.4464  & 0.0000 \\
$\sigma_{e,1}^2$  & 0.0329  & 0.0000 & 12937.0922 & 0.0000 \\
$\sigma_{e,2}^2$  & 0.0472  & 0.0000 & 9120.1400  & 0.0000 \\
$\sigma_{2,12}$   & 0.0157  & 0.0000 & 7473.0427  & 0.0000 \\ \hline
\end{tabular}
\end{table}

\bibliographystyle{apalike}
\bibliography{reference}

\end{document}